\documentclass[fleqn,twoside]{article}
\usepackage{amssymb}
\usepackage{espcrc2}

\usepackage{graphicx}
\usepackage[figuresright]{rotating}
\usepackage{epsfig}
\newcommand  {\vev}[1]{\left\langle #1 \right\rangle}

\newcommand{\AmS}{{\protect\the\textfont2
  A\kern-.1667em\lower.5ex\hbox{M}\kern-.125emS}}
\newcommand{\beq}{\begin{equation}}
\newcommand{\eeq}{\end{equation}}
\newcommand {\beqa}{\begin{eqnarray}}
\newcommand {\eeqa}{\end{eqnarray}}

\newcommand {\Tr}{\mbox{Tr\,}}

\newcommand {\dd}{\mbox{d}}
\newcommand {\ee}{\mbox{e}}

\newcommand  {\rf} [1]{(\ref{#1})}
\newcommand {\new}{{\rm new}}
\font\mybbs=msbm10 at 9pt
\def\bbs#1{\hbox{\mybbs#1}}

\def\IRs{{\bbs R}}

\title{On the Quantum Geometry of String Theory}

\author{J. Ambj\o rn\address[NBI]{The Niels Bohr Institute, Blegdamsvej
        17, DK-2100 Copenhagen \O, Denmark},
        K. N. Anagnostopoulos \address{Department of Physics,
        University of Crete, GR-710 03 Heraklion, Greece},
        W. Bietenholz \address{Institut f\"{u}r Physik,
Humboldt Universit\"{a}t zu Berlin, 
Invalidenstr. 110, D-10115 Berlin, Germany},
        F. Hofheinz\addressmark\ 
        and 
        J. Nishimura\addressmark[NBI]}
       
\begin{document}

\begin{abstract}
The IKKT or IIB  matrix model has been proposed as a non-perturbative
definition of type IIB superstring theories.
It has the attractive feature that space--time
appears dynamically. It is possible that lower dimensional universes
dominate the theory, therefore providing a dynamical solution to the
reduction of space--time dimensionality. We summarize recent works
that show the central role of the phase of the fermion
determinant in the possible realization of such a scenario. 
\vspace{1pc}
\end{abstract}

\maketitle

One of the most important problems in string theory is the emergence
of four dimensional space--time in low energy
physics. Compactification schemes have been the most popular approach
of the issue, but the price to pay is loss of predictability.  One
possible scenario is that 4d space-time appears dynamically from an
originally higher dimensional theory, usually $10D$ or $11D$. A
promising model that might realize this scenario is the IKKT matrix
model \cite{IKKT}, which is supposed to  define type IIB superstrings
in the large $N$ limit non-perturbatively.  One can view this model as
a ``lattice string 
theory'' and study it using standard Monte Carlo techniques.  Formally
it is the zero-volume limit of $D=10$, ${\cal N}=1$ super Yang-Mills
theory.  The partition function of the model (and its generalizations
to $D=4,6$) can be written as
\beq
Z =  \int \dd A ~ \ee^{-S_{\rm b} }~Z_{\rm f} [A]   \ ,
\label{original_model}
\eeq
where $A_{\mu}$ ($\mu = 1,\cdots , D$) are $D$ bosonic $N \times N$
traceless hermitian matrices, and $S_{\rm b}= -
\frac{1}{4g^2}\Tr([A_{\mu},A_{\nu}]^2)$ is the bosonic part of the
action.  The factor $Z_{\rm f} [A]$ represents the quantity obtained
by integration over the fermionic matrices, and its explicit form is
given for example in Ref.\ \cite{AABHN}. Space--time appears
dynamically (the eigenvalues of the bosonic matrices $A_\mu$ represent
space--time points) and if the dominant configurations have $d$
extended dimensions and $D-d$ small ones then we obtain essentially a
$d$ dimensional space--time {\it dynamically}. This requires in
particular the spontaneous symmetry breaking (SSB) of the manifest
$SO(D)$ invariance of the model. We stress here that the parameter $g$
appearing in eq.\ \rf{original_model} is not a coupling constant but a
scale which can be absorbed by a redefinition of the fields.  One
hopes to arrive at $d=4$ in the $10D$ IKKT model.

Such a scenario has not been verified. There are severe technical
problems in studying the $10D$ IKKT model by computer simulations,
related to the complex action problem. The study of simpler models has
shed some light on the possible mechanisms that could realize the
above mentioned scenario. We have learned that SSB is not realized in
the absence of the rapidly oscillating phase $\Gamma$ of the (complex)
fermionic partition function $Z_{\rm f} [A]$. In Ref.\ \cite{HNT} it
has been shown that when fermions are absent (``bosonic model'') SSB
does not occur. The study of a low energy approximation of the $10D$
and $6D$ models, where $\Gamma$ is set to zero by hand, did not reveal
any indication for SSB \cite{branched}. When the effect of $\Gamma$ is
infinitely enhanced in the $10D$ case by appropriately deforming the
original model, it turned out that space time is $2<d\le 8$, i.e. the
SSB scenario {\it is} realized \cite{NV}.  Recently it has been shown
in exactly solvable matrix models that SSB of $SO(D)$ occurs precisely
due to the phase of the fermion determinant \cite{JN}.

The purpose of this work is twofold: First to show that in the $4D$
model SSB does not occur \cite{4dSSB}. The $4D$ model has the property
that $Z_{\rm f} [A]$ is real positive, $\Gamma=0$, and it can be
studied using ordinary Monte Carlo methods. Therefore the result
supports the importance of the phase in the realization of SSB. The
authors of Ref.\ \cite{Burda:2000mn} have raised the possibility that
one dimensional structures dominate eq.\ \rf{original_model}. We
propose different, more physical probes of space--time dimensionality
that need to be used in $D=4$. Due to the power tail of eigenvalue
distributions \cite{KNS}, the observable $T_{\mu\nu} = \frac{1}{N} \Tr
(A_\mu A_\nu)$ used as a probe in Ref.\ \cite{Burda:2000mn} is ill
defined (note that $\vev{\Tr A^2}$ is divergent).  This does not
happen in higher dimensions where one expects all probes discussed
here to give identical answers. Of course in any dimension $\vev{\Tr
A^m}$ diverges for large enough $m$, therefore it is important to
understand and resolve this puzzle in a convincing way.

Second we show that it is possible to attack the problem of SSB in
higher dimensions despite the complex action problem. We propose a
method that allows us to compute the $T_{\mu\nu}$ eigenvalue
distribution in $6D$ for large $N$. The method does not suffer from
the complex action problem and it might be applicable  to other 
models as well. The resulting distribution is qualitatively
different from the case $\Gamma =0$, due to the presence of
the phase. A double peak structure appears for
large enough $N$, raising the possibility that small eigenvalues
dominate for some dimensions while large ones dominate for others,
which would mean that SSB occurs. A more involved study, possible
however, is needed in order to resolve the issue.

%
A natural probe of space--time dimensionality is its ``moment
of inertia'' tensor $T_{\mu\nu} = \frac{1}{N} \Tr (A_\mu A_\nu)$. Its
$D$ eigenvalues $\lambda _1 > \lambda _2 > \dots > \lambda _{D} > 0$
represent the principal moments of inertia. In $4D$, however, its
largest eigenvalue is known to diverge \cite{4dSSB}, in agreement with
the divergence of $\vev{\Tr A^2}$ \cite{KNS}. In this case it
is possible to modify its definition to 
$$
 T ^{(\new)}_{\mu\nu} =
\left\langle \sum_{i < j} \frac{2(x_{i\mu} - x_{j\mu}
) (x_{i\nu} - x_{j\nu} )} {N(N-1)\sqrt{(x_i - x_j)^2}} \right\rangle \, .
$$
$x_{i\mu}$ are the space--time points defined
to be the diagonal elements of the bosonic matrices $A_\mu$ when they
are transformed into a form as close to simultaneously diagonalized as
possible \cite{HNT,AABHN}.
\vskip 6mm 
  \begin{center}
 {\epsfig{file=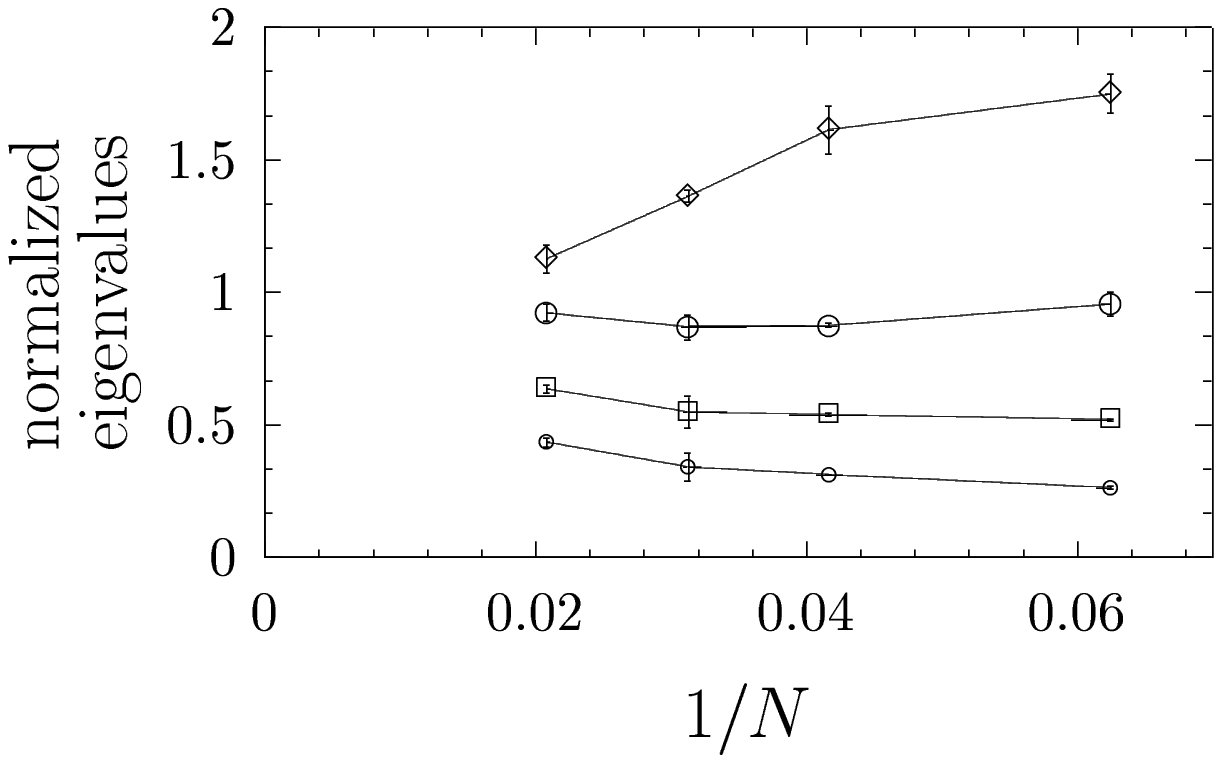,width=75mm}}
\end{center}
{\bf Fig. 1}: {\it The eigenvalues $\lambda^{\rm (new)}_i /(\sqrt{g}
  N^{1/4})$.} 
\vskip 6mm

  \begin{center}

 {\epsfig{file=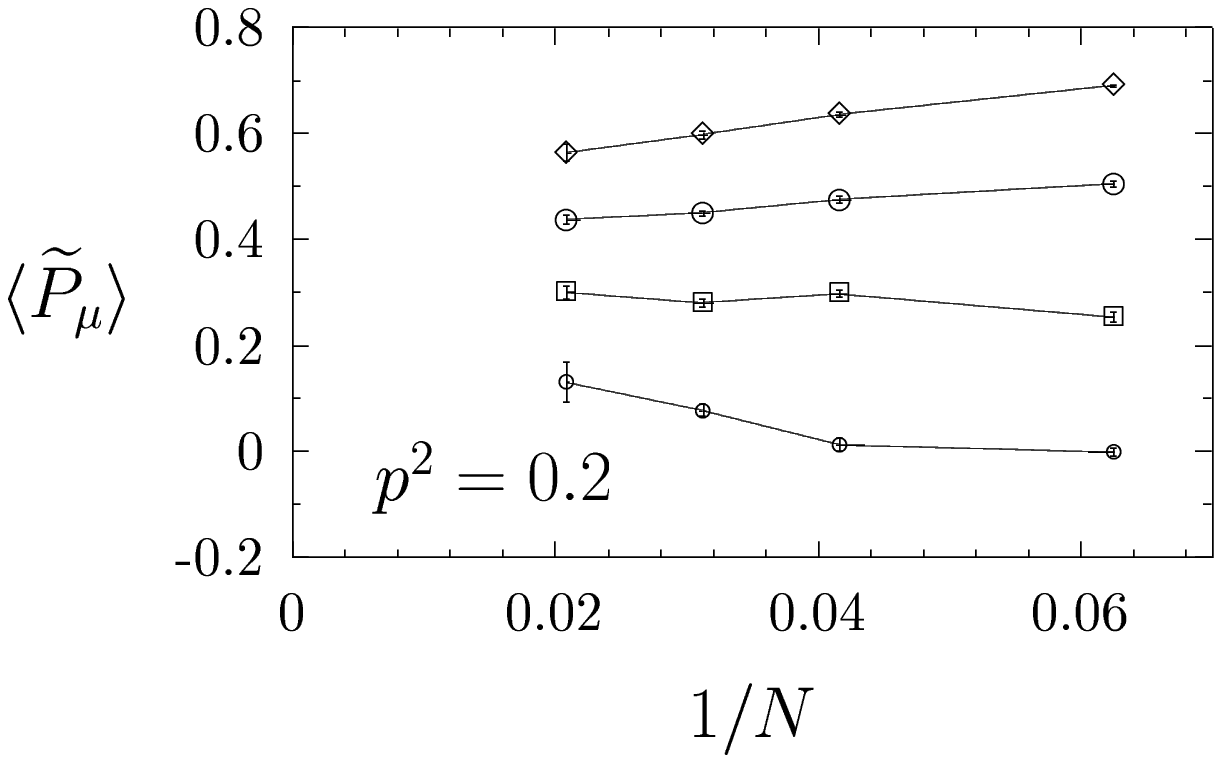,width=75mm}}
\end{center}
{\bf Fig. 2}: {\it The four Polyakov lines $\langle \tilde P_{\mu} (p)
\rangle$ 
($\mu = 1 , \dots , 4$) at $p^2 = 0.2$.}
\vskip 6mm
Since the distance between space--time points behaves asymptotically
as $\rho(r)\sim r^{-3}$, this quantity is well
defined. In Ref.\ \cite{4dSSB} we found that for increasing $N$, the
eigenvalues $\lambda^{(\new)}_1\dots\lambda^{(\new)}_4$ show no trend
for SSB. This can be seen in Fig. 1 where all eigenvalues seem to
approach the same value with increasing $N$. The same result is
obtained from 
the study of Polyakov lines $ P (\vec{p}) = \frac{1}{N} \Tr \, \exp (i
\, p_\mu A_\mu)$. They are essentially the Fourier transform of
$\rho(r)$ and their characteristic fall off is a measure of the extent
of space--time \cite{AABHN}. Since Wilson loops in general have been
interpreted as string creation operators they correspond to observables in
string theory. Defining the Polyakov lines on field configurations
$\tilde{A} _\mu$ that diagonalize $T_{\mu\nu}$ as $\tilde{P} _\mu (p)
= \frac{1}{N} \, \Tr \, \exp (i \, p \tilde{A} _\mu)$ we can study
if SSB occurs. In Fig. 2 we see that all $\tilde{P} _\mu (p)$
seem to converge to the same universal function, so no trend for SSB is
observed \cite{4dSSB}. This behavior is typical for all values
of $p^2$ that we measured.

For the $6D$ partition function eq.\ \rf{original_model} $\Gamma\ne 0$ and
the model suffers from the complex action problem. The partition
function can be written as
$
Z =  \int \dd A ~ \ee^{-S_0}~\ee^{i \Gamma}  \ ,
$
where $Z_{\rm f} [A] = \exp (\Gamma_{\rm R} + i \Gamma )$,
$\Gamma_{\rm R}, \Gamma \in \IRs $  and 
$S_0 = S_{\rm b} - \Gamma_{\rm R}$. The rapidly fluctuating
${\rm e}^{i\Gamma}$ 
term makes the calculation of expectation values exponentially hard
with increasing system size. Furthermore, the model defined by the
partition function  
$Z_0  =  \int \dd A ~ \ee^{-S_0 }$ used in the Monte Carlo simulation
visits very rarely the part of configuration space that dominates in the
full model (overlap problem). It is possible though to overcome
these technical obstacles. One can study the distribution of
$\tilde{\lambda} _i =\frac{\lambda _i}{\langle \lambda_i \rangle_0}$
defined by  
$
\rho_i (x) 
= \langle \delta (x - \tilde{\lambda} _i) \rangle
$, where $\vev{\ldots}_0$ are expectation values with respect to $Z_0$. In
Ref.\ \cite{AN}
we observed
\beq
\label{factrho}
\rho_i(x) = \frac{1}{C}\, \rho_i^{(0)}(x)\, \ee^{N^2 \Phi_i(x)}
\eeq
where $\rho_i^{(0)}(x)$ is the $\tilde{\lambda} _i$ distribution in the
$Z_0$ ensemble, 
$\Phi_i(x)=\frac{1}{N^2}\log\vev{\cos\Gamma}_{0,\tilde\lambda_i=x}$
and $C=\vev{\cos\Gamma}_{0}$ is a normalization constant. 
$\vev{\ldots}_{0,\tilde\lambda_i=x}$ are expectation values in the 
$Z_ {0,\tilde{\lambda}_i = x} = \int \dd A \, 
\ee ^{-S_0 } \, \delta (x - \tilde{\lambda}_i)$ ensemble. It turns out
that the function $\Phi_i(x)$ converges to an $N$--independent scaling
function for quite small $N$ ($\le 20$). Combining this observation with
the factorization property \rf{factrho}, we see that $\rho_i(x)$ can be
computed for much larger $N$ than $\Phi_i(x)$ since the
remaining terms in eq.\ \rf{factrho} do not suffer from the complex action
problem. 

In Ref.\ \cite{AN} we simulated the low energy version of the $6D$ IKKT
model \cite{IKKT} for $N$ up to $128$. We computed the scaling
function $\Phi_i(x)$, $i=4,5$  using $N\le 20$. 
We found that $\rho_i(x)$,  $i=4,5$ have two peaks at $x=x_<$ and
$x=x_>$ ($x_< < 1$ and $x_> > 1$), which is 
qualitatively different from $\rho^{(0)}_i(x)$, with peaks
around $1$. It is important to determine the peak which dominates
the large--$N$ limit. In Ref.\ \cite{AN} we found that the
data are consistent with a scenario where $x_<$ dominates $\rho_5(x)$
whereas $x_>$ dominates $\rho_4(x)$. This behavior can
be expected since the qualitative difference in the behaviors of
$\rho_5(x)$ and  $\rho_4(x)$ is mainly due to
$\rho^{(0)}_5(x)$ and  $\rho^{(0)}_4(x)$. In the branched polymer
description of the low energy effective theory of the IIB matrix model
\cite{IKKT},  $\rho^{(0)}_4(x)$ is expected to be much more suppressed
in the small $x$ regime than $\rho^{(0)}_5(x)$.
The method described
above is quite general and can be applied to other physical systems
that suffer from the complex action problem. Such a system is finite
density QCD, and we are currently investigating the related random
matrix theory with very encouraging results \cite{AANV}.

\end{document}